\documentclass[preprint]{ptephy_v1}
\usepackage[latin1]{inputenc}
\usepackage[T1]{fontenc}
\usepackage[english]{babel}
\usepackage{amsmath}
\usepackage{amssymb,amsfonts,textcomp}
\usepackage{color}
\usepackage{array}
\usepackage{supertabular}
\usepackage{hhline}
\usepackage{hyperref}
\newcommand\textstyleStrong[1]{\textbf{#1}}
\newcommand\textstylest[1]{#1}
\makeatletter
\newcommand\arraybslash{\let\\\@arraycr}
\newcommand\normalsubformula[1]
{\text{\mathversion{normal}$#1$}}
\begin{document}

\title{Adsorption of atoms and molecules on $s$-triazine sheet with embedded manganese atom: First-principles calculations}

\author{Yusuf Zuntu Abdullahi}
\affil{School of Physics, Universiti Sains Malaysia, 11800 Penang, Malaysia}
\affil{Department of Physics, Faculty of Science, Kaduna State University, P.M.B. 2339 Kaduna State, Nigeria 
 \email{yusufzuntu@gmail.com}}

\author{Tiem Leong Yoon}
\affil{School of Physics, Universiti Sains Malaysia, 11800 Penang, Malaysia
\email{tlyoon@usm.my}}

\author{Mohd Mahadi Halim}
\affil{School of Physics, Universiti Sains Malaysia, 11800 Penang, Malaysia}

\author{Md. Roslan Hashim}
\affil{Institute of Nano-Optoelectronics Research and Technology Laboratory, Universiti Sains Malaysia, 11900 Penang, Malaysia}

\author{Thong Leng Lim}
\affil{Faculty of Engineering and Technology, Multimedia University, Jalan Ayer Keroh Lama, 75450 Melaka, Malaysia}

\author{Kazuhiko Uebayashi}
\affil{Department of Natural Science, Akita College, National Institute of Technology,
Akita 011-8511, Japan}

\begin{abstract}
The mechanical, structural, electronic and magnetic properties of \textit{s}{}-triazine sheet
(C\textsubscript{6}N\textsubscript{6}) with embedded Mn atom (Mn-C\textsubscript{6}N\textsubscript{6}) is investigated
under the influence of external environment using density functional theory. Our results show that
Mn-C\textsubscript{6}N\textsubscript{6} system is structurally and mechanically stable. The binding energy of Mn
embedded in C\textsubscript{6}N\textsubscript{6} sheet can be modulated under the influence of symmetric deformation
and perpendicular electric field respectively. The semiconducting property of pure C\textsubscript{6}N\textsubscript{6}
sheet is maintained upon embedment of Mn atom in the porous site. It is also found that small increment in bi-axial
tensile strain enhances the band gap (from 0.630 eV at zero strain to 0.802 eV at 5\% strain) while the magnetic moment
of the embedded Mn atom is preserved. The electronic and magnetic properties of the
Mn-C\textsubscript{6}N\textsubscript{6} systems are maintained up to 10 V/nm in electric field strength. We also
explore the geometries, electronic and magnetic properties of Mn-C\textsubscript{6}N\textsubscript{6} with adsorbed
atoms and molecules. The Mn-C\textsubscript{6}N\textsubscript{6} with adsorbed O atom and O\textsubscript{2} molecule
systems shows half-metallic character whereas the remaining systems preserve their semiconducting property. The total
magnetic moment per unit cell in most of the systems is found to reduce as compared to that of the
Mn-C\textsubscript{6}N\textsubscript{6} sheet. The reduction in magnetic moment can be related to the strong
interactions among the Mn atom and the surrounding atoms which lead to the formation of low-spin configurations.
Overall, our results indicate that the Mn-C\textsubscript{6}N\textsubscript{6} systems with and without adsorbed atoms and molecules can serve as potential candidates for future spintronics and catalysis applications.
\end{abstract}


\maketitle

\section{Introduction}

Research interest in magnetic materials has gained its spectacular boost after the discovery of exotic magnetic
properties [1, 2]. Additionally, continued device miniaturization has also opened up many research opportunities to
reveal the magnetic properties of these nanostructures in hope for potential applications in spintronics and catalysis.
At present, investigations on magnetic nanostructure are directed towards clusters of transition metals (TM). This is
due to their large magnetic moments as the dimensionality and coordination (bonds per atom) of the TM cluster are
reduced [3, 4]. However, their net magnetic moments can also become decreased, depending on the configuration
(ferromagnetic and anti-ferromagnetic geometries) and surrounding of the clusters [5-8]. It is desirable to study the
magnetic properties of a TM cluster which is evenly bound to an inert substrate such that the magnetic moment is
comparable to that of a cluster in isolation. The interaction of TM atoms with certain metal surfaces is known to be
very strong [8,9]. This could potentially affect their net magnetic moments. In contrast, carbon nanostructures are
known to be inert, and stand as the right candidate for encapsulation of TM atoms.

{ 
Considering two-dimensional carbon nanostructures as substrates, many efforts have been devoted to study the physical
and chemical properties of bound TM clusters on graphene [6,7,10] and boron nitride [11,12] sheets. However, previous
works have shown that TM atoms bind weakly on graphene and boron nitride monolayers [7,12,13]. Such a weak binding may
render TM atoms to diffuse easily on the surface and aggregate to form clusters. Many attempts have been made to create
defects in the aforementioned surfaces to prevent the diffusion. However, not much success [12] has been achieved due
to technical impracticability to realize controlled defect formation in compacted hexagonal 2D surfaces. Binding of
dispersed TM atoms on the surfaces of graphene and other related 2D monolayers presents a practical challenge. }

{ 
To practically produce controllable nanostructures with desired magnetic properties without altering the physical
properties of the bound structures is not technically trivial. Such a challenge has led to the discovery of a new set
of graphitic carbon-nitride (CN) sheets [14]. Graphitic CN sheet comes naturally with a regular array of cavities which
is wide enough to accommodate any nanostructures that are evenly dispersed. CN sheet has many graphitic allotropes
depending on the choice of the primitive unit cell and the C and N atomic coverage in their hexagonal rings. For
example, the semiconducting heptazine (\textit{t}{}-\textit{s}{}-triazine) sheet comprises of 14 C and N atoms in the
form of three abreast rings of triazine (consisting of 3 C and 4 N atoms in the unit cell) [15] whereas
\textit{s}{}-triazine sheet of similar electronic character consists of 12 C and N atoms having two rings of triazines
linked together by C-C bond [16]. Both aforementioned allotropes are found to be structurally, energetically and
mechanically stable under ambient conditions [15-17]. }

{ 
Having explained the motivation for the choice of graphitic CN sheet over the rest of 2D sheets with compacted hexagonal
rings, the embedment of single atoms on the pure graphitic CN sheet has been the most recent interest for many
theoretical [18-21] and few experimental [22-25] studies.\textcolor{black}{ }The structural, electronic and magnetic
properties of B, Al, and Cu atoms doped graphitic triazine-based
\textit{g}{}-C\textsubscript{3}N\textsubscript{4}~(\textit{gt}{}-C\textsubscript{3}N\textsubscript{4}) systems have
been theoretically investigated [26]. Their results have shown that Cu atoms are more stable when situated above the
center of the triazine ring. Moreover, they have reported that doping at the interstitial sites leads to
thermodynamically stable non-planar structures. It is found that Cu-doped triazine produces a total magnetic moment of
1.0~\textit{$\mu $}B~which is mainly localized around  $p_z$ like-orbitals of the sheet. The electronic property of
Cu-doped triazine systems calculated based on GGA-PBE and HSE06 method exhibits half-metallic behavior. The Cu-doped
triazine system has an anti-ferromagnetic ground state. They did not estimate the Curie temperature for Cu-doped
triazine system due to the fact that long-range magnetic ordering is usually favorable for the system having
ferromagnetic order at zero temperature.}

{ 
The energetic, electronic, magnetic and optical properties of two-dimensional graphitic carbon nitride
(\textit{g}{}-C\textsubscript{3}N\textsubscript{4}) sheets upon embedment of 3d-transition metal (TM) atoms have been
investigated by Ghosh \textit{et al}. The \textit{g}{}-C\textsubscript{3}N\textsubscript{4}~sheets with embedded TM
atoms are found to be energetically stable when located in the porous sites. It was also reported that the system
becomes metallic upon TM (including Cu) embedment. The d orbitals of TMs hybridize with the $\pi $-orbitals of the
\textit{g}{}-C\textsubscript{3}N\textsubscript{4}~framework and close the band gap in TM-embedded
\textit{g}{}-C\textsubscript{3}N\textsubscript{4}~(TM-\textit{g}{}-C\textsubscript{3}N\textsubscript{4}). In most cases
the magnetic moment of the 3d TM atoms are well preserved when compared to their isolated values. However, Mn atoms
couples antiferromagnetically whereas Cu and Zn atoms are nonmagnetic in the ground state of their corresponding
TM-\textit{g}{}-C\textsubscript{3}N\textsubscript{4}~sheets [21]. The claims for magnetic ordering in Mn, Cu embedded
\textit{g}{}-C\textsubscript{3}N\textsubscript{4} sheets and metallic behavior for Cu embedded system at relatively the
same separations are in contrast to what was observed by Zhang \textit{et al.} [20] and Meng \textit{et al.} [26] in
the same allotropes. In a similar allotrope, Du \textit{et al.} [27] suggest that Mn and Cu atom embedded in
C\textsubscript{2}N monolayers possess ferromagnetic and paramagnetic ground state respectively when the TM atoms are
close to each other, while Meng \textit{et al}. [26] observed antiferromagnetic ground state with half-metallic
electronic character for Cu-doped triazine system. In general, the claim for magnetic ordering of TM atoms embedded in
CN remains unclear. However, it is feasible to envisage that if the embedded TM atoms were well separated, the possible
interactions among themselves could be minimized. The effort to clarify the magnetic ordering issue hence could become
less complicated. }

{ 
More recently, Choudhuri \textit{et al.} [27] investigated newly formed allotrope of the carbon-nitride monolayer
(\textit{gt}{}-C\textsubscript{3}N\textsubscript{3}) with embedded 3d transition metal atoms
(TM@gt-C\textsubscript{3}N\textsubscript{3}) systems based on density functional theory. It was found that the Cr, Mn
and Fe embedded \textit{gt}{}-C\textsubscript{3}N\textsubscript{3}~systems are dynamical, thermally and mechanically
stable. Moreover, Cr, Mn and Fe embedded in the \textit{gt}{}-C\textsubscript{3}N\textsubscript{3}~systems have been
found to exhibit high temperature ferromagnetism and high magnetic anisotropy energy (MAE) with a peak value per atom
occurring in Cr@\textit{gt}{}-C\textsubscript{3}N\textsubscript{3}. Their finding for the ferromagnetic ordering of Mn
atom in similar graphitic CN sheet under ambient conditions agrees well with that reported by Du \textit{et al.} [28].
Choudhuri \textit{et al.} [27] further claimed a more enhanced MAE in the presence of an external electric field, an
amount far larger than the value computed without electric field.}

{ 
In our recent studies, it was shown that the band gap of heptazine with embedded Mn atom increases under small biaxial
tensile strain whereas the magnetic moment of heptazine with embedded Mn systems remain unperturbed under both tensile
strain and applied electric field [18]. It was also observed that the binding energy of heptazine with embedded Mn atom
systems decreases as more tensile strain is applied. In contrast, we observed significant increase in binding energy as
the strength of electric filed is increased. Modulating binding energy via an applied external field could serve as an
effect way to control catalytic process.}

{ 
Besides spintronics, CN with embedded TM systems can also be utilized for heterogeneous catalysis and in membrane for
hydrogen purification from other gases [29-31]. Heterogeneous catalysis based on single atomic crystal (SAC) is the
cheapest and easiest way of achieving efficient catalysis. To date, various 2D sheets have been theoretically and
experimentally proposed for heterogeneous catalysis [32-37]. Among others, monolayers carbon and related 2D
nanostructures, e.g. boron nitride [32,33], graphene [34,35], graphyne [36], silicene [37], carbon nitride sheet [35]
have received enormous attentions due to their availability, thermal stability and large surface area. Density
functional theory (DFT) calculations have shown that graphene with embedded Au, Fe and Pt [34,38,39] atoms yield good
catalytic activity for CO oxidation and oxygen reduction. Experimentally, excellent catalytic activities have been
reported for graphene with supported single Pd and Pt using monolayer deposition method [29,40]. Although favorable
catalytic process has been reported for SACs of graphene and related 2D materials, the adsorption/binding energies of
3d atoms on their surfaces are relatively weak and could result in formation of cluster that renders reduction in
catalytic activity. }

{ 
2D porous carbon sheets embedded with TM atoms [14,16] are desirable SACs that can serve as membrane for hydrogen
purification from other gases [30,31]. This is because they naturally come with regularly array of cavities. These
cavities are reported to strongly bind to 3d TM while maintaining structural and electronic features of the sheets
[18,21]. A recent work along this direction has confirmed theoretically that it is possible to synthesize high quality
SACs using porous 2D materials with well dispersed 3d TM atoms. More importantly, the 3d TM atoms are found to strongly
bind to the C\textsubscript{2}N sheet in the cavity without cluster formation, while the intrinsic properties of both
porous substrate and the atoms are preserved [41]. }

{ 
In this paper, we are interested to theoretically study the geometries, electronic and magnetic properties of
TM-embedded graphitic systems in which different types of atoms and molecules are made to adsorbed onto their surfaces.
To be specific, Mn-embedded \textit{s}{}-triazine will be studied, as well as the adsorption of C, N, O, H atoms and
six molecules (CH\textsubscript{4}, N\textsubscript{2}, O\textsubscript{2}, H\textsubscript{2}, CO,
CO\textsubscript{2}) on its surface. \textit{s}{}-triazine sheet is an allotrope of graphitic carbon nitride which
shares some structural similarities with C\textsubscript{2}N sheet. It also comes with well separated porous sites
which will ensure no interactions between the embedded TM atoms. Mn is chosen instead of other 3d TM species due to its
well-ordered spin configurations in the d orbital and desirable characteristics for SAC applications which was reported
in previous works. }

\section{Calculation method}
We carried out density functional theory (DFT) [42] calculations using the QUANTUM ESPRESSO code [43]. Generalized
gradient approximation (GGA) of Perdew-Burkew-Enzerhof (PBE) [44] and Hubbard \textit{U} correction [45]
(GGA+\textit{U}) is employed to describe the exchange-correlation energy of strongly localized 3d orbital of Mn atom.
Ultrasoft pseudopotentials are used for C, N, and Mn (semi-core included) atoms to account for the core and valence
electrons [46]. We employed plain wave basis set with kinetic energy cut-off of 550 eV to expand the wave functions.
Marzari-Vanderbilt smearing method with Gaussian spreading is employed [47] to facilitate the convergences of our
systems. We used 8{\texttimes}8{\texttimes}1 and 15{\texttimes}15{\texttimes}1 k-point meshes of Brillouin zone (BZ)
sampling using the Mankhorst-Pack method for the self-consistent field and density of state calculations respectively
[48].

\begin{figure}[!p]
\begin{center}
 \includegraphics[width=4.2in,height=1.7in]{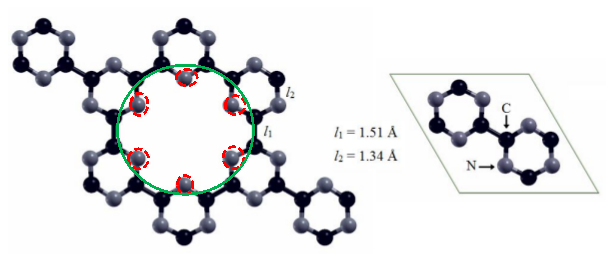} 
 \includegraphics[width=5.5in,height=2.5311in]{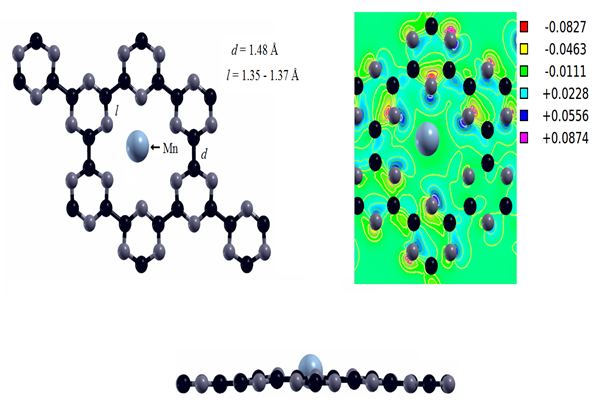} 
\end{center}
 \caption{
(a) {\bf Left}: Top view of the optimized 2{\texttimes}2 geometry of
C\textsubscript{6}N\textsubscript{6}. The atomic symbols and the calculated bond leghts are shown in the figure. Atoms
in dotted red circles are the edge nitrogen atoms (\textit{N}\textsubscript{edge}). There are
\textit{N}\textsubscript{edge}=6 of them around the cavity. \textbf{Right}: unit cell of \textit{s}{}-triazine which is
made of 2 C\textsubscript{3}N\textsubscript{3} rings. 
(b). Left: Top view of a supercell comprised of 2{\texttimes}2 \textit{s}{}-triazine unit cell, with an
embedded Mn atom (Mn-C\textsubscript{6}N\textsubscript{6}). This Mn-C\textsubscript{6}N\textsubscript{6} is the one
that was used as the input in our calculation. \textbf{Right}: Top view plot of the charge-density difference of the Mn
atom and the surrounding C and N atoms. The color scale shows ranges of charge accumulation and depletion in a.u.
Carbon atoms are in black. Atoms dotted with 4-point starts are the \textit{N}\textsubscript{edge} atoms (nitrogen
atoms in dark ash color). \textbf{Down}: Side view of the optimized buckled Mn-C\textsubscript{6}N\textsubscript{6}
under applied perpendicular electric field.
}
\label{fig:fig1}
\end{figure}

\begin{figure}[!p]
\begin{center}
(a) \includegraphics[width=2.86in,height=2.618in]{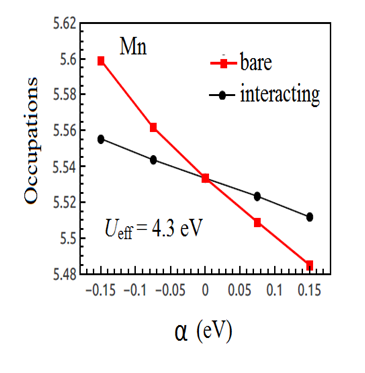} 
(b)\includegraphics[width=2.59936in,height=2.38in]{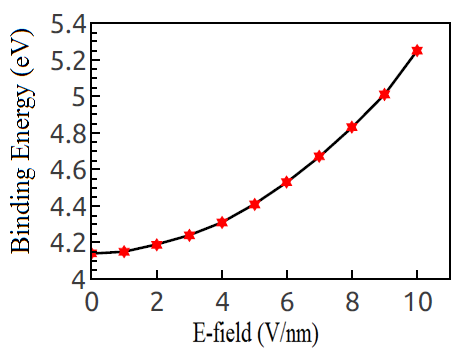} \\
(c) \includegraphics[width=2.59936in,height=2.38in]{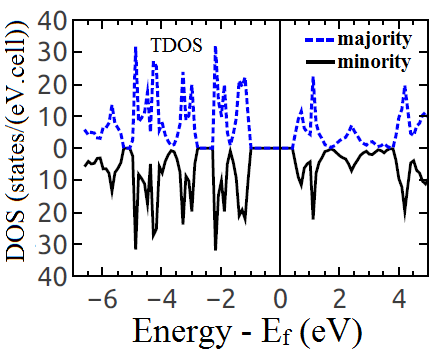} 
(d) \includegraphics[width=2.59936in,height=2.38in]{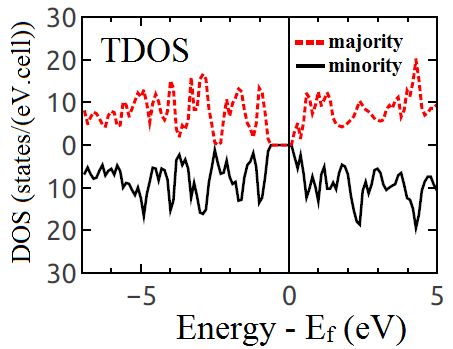}
\end{center}
\caption{
(a) {\bf Left}: Linear response of d orbital occupations as a function of potential shift  $\alpha $. The
curves depicted by the dotted red and black lines are labeled bare and interacting. The inverse response functions are
deduced numerically by calculating the slope of the curves.  $\chi _0$ follows from the slope of curve bare, whereas 
$\chi $\textcolor{black}{\textsubscript{ }}from the slope of curve interacting. \textbf{Right}: The magnetic moment per
unit cell and binding energy depicted by the dotted red and black lines respectively as a function of an applied
electric field for Mn-C\textsubscript{6}N\textsubscript{6} system. \textbf{Down}: The spin-polarized total density of
state (TDOS) for Mn-C\textsubscript{6}N\textsubscript{6} under applied electric field.}
\label{fig:fig2}
\end{figure}

{ 
The relaxed structure of 2{\texttimes}2 pure graphitic C\textsubscript{6}N\textsubscript{6} sheet (left figure in Fig.
~\ref{fig:fig2}(a)) consists of 4 units of \textit{s}{}-triazine (right figure in Fig.~\ref{fig:fig1}(a)) with a total of 48 C and N atoms. Each
\textit{s}{}-triazine has two rings of C\textsubscript{3}N\textsubscript{3} connected via a C-C bond. The averaged
optimized C-C bond length connecting adjacent \textit{s}{}-triazine unit, C-N bond length and the lattice constant in
the pure C\textsubscript{6}N\textsubscript{6} are found to be 1.51 \AA\textit{ }, 1.34 \AA\textit{ } and 14.27 \AA\textit{ } respectively (see the left
figure in Fig.~\ref{fig:fig1}(a)). In the \textit{s}{}-triazine structure, all C atoms are sp\textsuperscript{2}{}-like hybridized
whereas N atoms have sp\textsuperscript{3}{}-like hybridized structure. Porous site in the graphitic
C\textsubscript{6}N\textsubscript{6} is encircled by solid line in the left figure in Fig.~\ref{fig:fig1}(a). The left figure in
Fig 1. (b) shows Mn atom embedded in the porous site. Henceforth, we shall refer the 2{\texttimes}2 supercell in Fig.
1(a) as C\textsubscript{6}N\textsubscript{6}, while Mn embedded in C\textsubscript{6}N\textsubscript{6} as
Mn-C\textsubscript{6}N\textsubscript{6} in the left figure of Fig.~\ref{fig:fig1}(b). A thick vacuum space of 16 \AA\textit{ } along the
direction perpendicular to the Mn-C\textsubscript{6}N\textsubscript{6} sheet is adopted to avoid the interaction
between the periodic sheets. All atoms are fully relaxed until all remaining force on each atom was smaller than 0.05
eV/\AA\textit{ }.}

{ 
As suggested in our previous works [18], inclusion of Hubbard \textit{U} parameter yield better ground state properties
for strongly correlated systems involving TM atoms. We therefore employed linear response approach formulated by
Cococcioni \textit{et al}. [45], which is implemented in the PWSCF package [43], to compute the \textit{U} parameter in
this study. In the linear-response scheme calculations, we first evaluated  $\chi _0$ and  $\chi $ which represent the
non-interacting (bare) and interacting density response functions of the system with respect to localized
perturbations, as illustrated in Fig.~\ref{fig:fig2}(a). The parameter \textit{U}\textsubscript{eff} is then obtained from the
expression:  $U_{\normalsubformula{\text{eff}}}=\left(\chi _0^{-1}-\chi ^{-1}\right)$ . The calculated
\textit{U}\textsubscript{eff} value for Mn-C\textsubscript{6}N\textsubscript{6} is depicted in Fig.~\ref{fig:fig2}(a). }

\section{Results ad Discussions}

\subsection{Structural and mechanical properties of Mn-C$_6$N$_6$ system}

\begin{figure}[!p]
\begin{center}
\includegraphics[width=4.5in,height=3.451in]{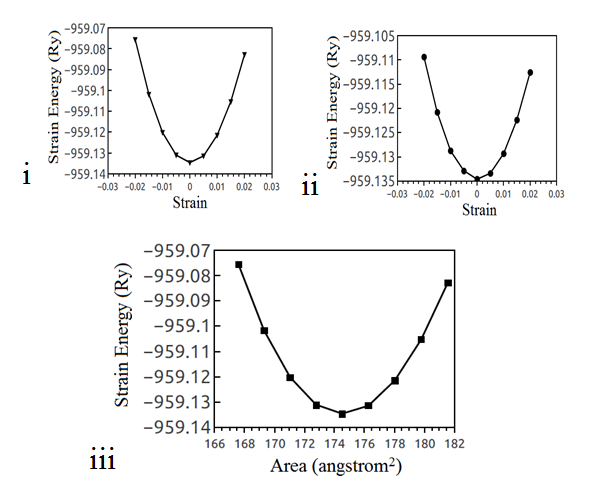}
\end{center}
\caption{Dependence of strain energy (Ry) on (i) bi-axial (ii) uni-axial and (iii) area of the Mn-C\textsubscript{6}N\textsubscript{6} system for elastic constant calculation.}
\label{fig3}
\end{figure}

{ 
Fig.~\ref{fig:fig1}(c) displays the relaxed structure of Mn-C\textsubscript{6}N\textsubscript{6}. The averaged optimized C-C bond
length connecting adjacent \textit{s}{}-triazine unit and the lattice constant in the
Mn-C\textsubscript{6}N\textsubscript{6} are found to be 1.48 \AA\textit{ }\textit{ }and 14.20 \AA\textit{ }\textbf{ }respectively. The optimized C-N
bond lengths within the porous site are in the range 1.35 - $\sim$1.37 \AA\textit{ }. The calculated values do not deviate much from
the un-doped C\textsubscript{6}N\textsubscript{6} system and is consistent with that of the previous work [16]. Using
strain energy minimization procedures in steps of 0.005 for both uni- and biaxial tensile strains within the linear
elastic region (see Fig.~\ref{fig3} and Table~\ref{table1}), mechanical properties are obtained. The elastic constants
\textit{k}\textsubscript{11}, \textit{k}\textsubscript{12}, for uni- and bi-axial strains are 146.90 N/m and 26.20 N/m
respectively, which are obtained from Eq.~\eqref{eq1}. The in-plane stiffness (Young modulus), and poison ration can be deduced
from Eq.~\eqref{eq2}. These equations are expressed as:}

\begin{eqnarray}
k_{11}&=&{1 \over A_0}\left[ {\partial^2 U \over \partial s^2}\right]_{s=0} \ \ \ (\textrm{uniaxial}) \nonumber \\
2(k_{11}+k_{12})&=&{1 \over A_0}\left[ {\partial^2 U \over \partial s^2}\right]_{s=0} \ \ \ (\textrm{bi-axial})
\label{eq1}
\end{eqnarray}
In-plane stiffness and Poisson's ratio:
\begin{eqnarray}
Y&=&k_{11}(1 - \nu^2), \nonumber \\
\nu &=& {k_{12} \over k_{11}}
\label{eq2}
\end{eqnarray}
where  $k_{11},k_{22}$ denote elastic constants,
\textit{A}\textsubscript{0}, \textit{U}, and \textit{s} are equilibrium unit-cell area, strain energy and applied
tensile strain. The computed in-plane stiffness and Poisson's ratio are 1421.4 GPa${\cdot}$\AA\textit{ } (= 142.14\textbf{ }N/m)
and 0.18 respectively. This shows that Mn-C\textsubscript{6}N\textsubscript{6} is less stiff than heptazine [15] and
graphyne sheets [48]. The decreased in stiffness in comparison with heptazine can be linked to a weaker bonding of the
hexagonal rings of the sheet. Moreover, the calculated Poisson's ratio is slightly larger than the value for graphene
[50]. To evaluate the bulk modulus, we use second derivative of the bi-axial strain energy as a function of area of the
Mn-C\textsubscript{6}N\textsubscript{6} sheet, which is expressed as
\begin{equation} \label{eq3}
G=A\times \left.\left(\frac{{\partial}^2 U}{{\partial}A^2}\right)\right|_{A_m}
\end{equation}
where \textit{A}, \textit{U} and \textit{A}\textsubscript{m} denote the area of the
Mn-C\textsubscript{6}N\textsubscript{6} sheet, the bi-axial strain energy and the equilibrium area of
Mn-C\textsubscript{6}N\textsubscript{6} respectively. Fig.~\ref{fig3} iii shows the bi-axial strain energy as a function of the
area of Mn-C\textsubscript{6}N\textsubscript{6} sheet. The computed value for the bulk modulus of
Mn-C\textsubscript{6}N\textsubscript{6} sheet is 86.49 N/m which is also lower than the value for Mn embedded in
heptazine sheet [18]. The lowering of the value for bulk modulus can also be related to the compact hexagonal rings
within the heptazine sheet as compared to separated rings in \textit{s}{}-triazine sheet. All the computed results of
elastic constants confirm that Mn-C\textsubscript{6}N\textsubscript{6} is mechanically stable.
{ 

To further examine the structural stability of Mn-C\textsubscript{6}N\textsubscript{6}, the averaged bond lengths
\textit{d}\textsubscript{Mn-N} between the embedded Mn and six nearest neighbor N atoms in the porous site, the
averaged bond lengths \textit{d} connecting adjacent \textit{s}{}-triazine, and \textit{h} the height of the Mn atom
with respect to the Mn-C\textsubscript{6}N\textsubscript{6} plane for strained and unstrained systems are listed in
Table~\ref{table2}. We have also checked the binding energy of Mn in the Mn-C\textsubscript{6}N\textsubscript{6} sheet as a
function of bi-axial tensile strain for the range of 0-6\%. The binding energy of the
Mn-C\textsubscript{6}N\textsubscript{6} is calculated as }
\begin{equation}
E_b=\left(E_{\mathrm{C}_6\mathrm{N}_6}+E_{\text{Mn}}\right)-E_T
\label{eq4}
\end{equation}
{ 
where \textit{E}\textit{\textsubscript{T}} is the total energy of Mn-C\textsubscript{6}N\textsubscript{6} sheet, 
$E_{\mathrm{C}_6 \mathrm{N}_6}$ represents the energy of pristine
C\textsubscript{6}N\textsubscript{6} sheet, and  $E_{\mathrm{Mn}}$ is the total energy of an isolated Mn atom. Positive
binding energy is an indication of a stable structure. According to Eq.~\eqref{eq4}, the unstrained
Mn-C\textsubscript{6}N\textsubscript{6} has a binding energy of 4.14 eV, which is roughly the same as that obtained in
the recent works [41,51]. As can be seen in Table~\ref{table2}, there is little difference in the predictions of structural and
magnetic properties of unstrained Mn-C\textsubscript{6}N\textsubscript{6} calculated in this work as compared to
previous results. Here we state clearly the differences in terms of methodology and the choice of the unit cell between
our calculations and the previous works. While we consider DFT+\textit{U} calculations in our
Mn-C\textsubscript{6}N\textsubscript{6} system, the work in Ref. [51] employed only DFT without Hubbard \textit{U} in
the presumably same system. We opine that accurate computation of geometries and electronic properties in a system
involving strongly correlated atoms can only be achieved if Hubbard \textit{U} is included [45]. The main difference
between the present work and that of Ref. [41] lies within the choice of the unit cell. We use allotrope of graphitic
carbon nitride (CN) sheet with equal number of C and N in all the hexagonal rings, with two rings connected via C-C
bond in the unit cell with no dangling bond, whereas Ref. [41] used different allotrope of graphitic CN sheet. We
anticipate that our results should agree with theirs since the environment where Mn atom is placed is probably the
same.}

{ 
We have further investigated the energetic, structural, electronic and magnetic properties under uniform tensile strain
from 0\% to 6\%. For increasing tensile strain, the calculated value of the Mn height \textit{h} in the optimized
Mn-C\textsubscript{6}N\textsubscript{6} plane confirms the planarity of the sheet, where only slight movement of the Mn
within the plane was found. We also observed that as the applied tensile strain increases the binding energy of Mn in
the porous site becomes less pronounced. The decrease of the binding energy can be understood as follows. In general,
elastic energy,  $U$, for small deformations is positive. It is proportional to the product of square of the strain and
the elastic module(s) related to biaxial tensile strain, which is (are) supposed to be positive as well. So for
positive elastic moduli, the total energy (which is negative in value) of the Mn-C\textsubscript{6}N\textsubscript{6}
system should increase under small strain. Consequently, the binding energy (which is positive in value), according to
the definition of Eq.~\eqref{eq4}, would become lower. We observe a uniform increase in the average C-C bond lengths connecting
the heptazine units. Additionally, it was found that the average bond lengths between the embedded Mn and six nearest
neighbor N atoms in the porous site increase accordingly as well. Overall, the decrease in the binding energy is due to
structural distortion. At a 6\% tensile strain, metastable structure is formed in which the binding energy,  $E_b,$
becomes negative (see Table~\ref{table2})}

\subsection{ Electronic and magnetic properties Mn-C{\textsubscript{6}}{N}{\textsubscript{6}}
}

Our calculations show that Mn embedded in C\textsubscript{6}N\textsubscript{6} maintains its isolated magnetic moment
for both strained and unstrained systems. This indicates that the magnetic moment is less sensitive to the applied
bi-axial tensile strain. The corresponding magnetic moments for various systems are listed in Table~\ref{table2}, and the magnetic
moment for unstrained Mn-C\textsubscript{6}N\textsubscript{6} system agrees well with the previous literatures [41,51].
We can as well observe from the magnetic moments per atom in Table~\ref{table2} that the magnetism of
Mn-C\textsubscript{6}N\textsubscript{6} systems is mainly contributed by the Mn atom. Lowdin's charge analysis shows
that electron redistribution of Mn orbitals are mainly from s, p orbitals of Mn whereas d orbital relatively retain its
unpaired electrons (high-spin configuration is maintained in each deformation It is found that charge transfer into the
C\textsubscript{6}N\textsubscript{6} sheet,  $Q$, which is calculated from Lowdin's charge analysis, is relatively
maintained in all deformations.

{ 
\ \ The coupling (i.e., ionic interaction between the Mn atom and C\textsubscript{6}N\textsubscript{6} sheet) does not
decrease the number of spin-polarized d orbital electrons of Mn atom. We also show the top view plot of the
charge-density difference between the Mn atom and the surrounding C and N atoms of strain-free system in the right
figure of Fig.~\ref{fig:fig1}(b). The figure indicates charge localization around the most electronegative N edge (dotted black
color) atoms in the porous site. The charge depletion between the atoms confirms the covalent property of the bonds
[52].}

{ 
The total density of state (TDOS) of pure C\textsubscript{6}N\textsubscript{6} is illustrated in Fig. ~\ref{fig:fig2}(b). The pure
C\textsubscript{6}N\textsubscript{6} is a direct band gap semiconductor with a band gap of approximately 1.53 eV. The
calculated bang gap tallies well with the previous work [16]. The embedment of Mn atom into the
C\textsubscript{6}N\textsubscript{6} induces magnetism for Mn-C\textsubscript{6}N\textsubscript{6} system while
maintaining the intrinsic electronic property of the C\textsubscript{6}N\textsubscript{6} sheet. This is evident by
comparing Fig.~\ref{fig:fig2}(c) (which shows the TDOS of the pure C\textsubscript{6}N\textsubscript{6} sheet) and Figs.~\ref{fig4} (a)-(c)
(which shows the spin-polarized TDOS for the strain-free Mn-C\textsubscript{6}N\textsubscript{6} system). The
asymmetric TDOS of Mn-C\textsubscript{6}N\textsubscript{6} system confirms the presence of magnetic moment in the
system.}

{ 
The spin-polarized projected density of state (PDOS) in Figs.~\ref{fig4} (d)i - (f)i, illustrate the atomic orbital contributions
of Mn-C\textsubscript{6}N\textsubscript{6} system to valence and conduction bands around Fermi level. The 6 edge N
atoms in the porous site formed bonds with the embedded Mn atom. In the plane of C\textsubscript{6}N\textsubscript{6},
the lone pairs of N atoms face the  $d_{\mathit{xy}}$ and  $d_{x^2-y^2}$ orbitals of Mn atom. Due to repulsive effect
between the  $p_x,p_y$ orbitals of N atom and the  $d_{\mathit{xy}}$,  $d_{x^2-y^2}$ orbitals of Mn atom, the  $p_z$
orbital will be at higher energy. Additional evidence can be referred to Fig.~\ref{fig4}(d) which shows that the bottom of the
conduction band is dominated by  $p_z${}-like orbitals of the six edge N atoms in both spin states, whereas the other
orbitals (e.g.,  $s,p_x,p_y$) are located in the valence band (which is at a lower energy). This is supported by the
measured EC values as shown in Table~\ref{table2}, where these values increase with increasing tensile strain. Moreover, the top
of valence band displays dominance by lone pairs of N atoms. The remnant  $d${}-like orbitals would be forced to the
lower energies. At approximately -6.5 eV and -8.0 eV the bonding orbitals are mainly due to sp-like orbitals of the
edge N atoms, and s-,  $p_x${}- and  $d${}-like orbitals of the Mn atom. Around -4.5 eV there is a hybridization which
is occupied by s-,  $d_{\mathit{zx}}${}- and  $d_{z^2}${}- like orbitals of the Mn atom, as well as sp-like orbitals of
the edge N atoms in the minority spin state. \ The sp-like orbitals of the Mn atom in both spin up and spin down
channels portray intra-orbitals electron transfer.}

{ 
We also observe band gap modulation of Mn-C\textsubscript{6}N\textsubscript{6} under uniform deformation. Small
increment in the band gap when bi-axial tensile strain is applied can be explained as follows. From the PDOS Figs. 4
(d)ii-(f)ii of strained Mn-C\textsubscript{6}N\textsubscript{6} sheet (3\% bi-axial tensile strain), we observe a
uniform shift of  $p_z${}-like states towards higher energy and backward shift of lone pairs of N edge atoms as larger
bi-axial tensile strain is applied. This results in band gap increase. We note that the symmetric deformation causes
the lone pairs to become misaligned, thereby causing  $p_z${}-like orbital to shift towards higher energy so as to
reduce the steric repulsions between the  $p_x$,  $p_y$ orbitals of N atom and the  $d_{\mathit{xy}},d_{x^2-y^2}$
orbitals of Mn atom.}

\subsection{Electric field effect}

{ 
To study more effects of external environment on the Mn-C\textsubscript{6}N\textsubscript{6} sheet, we have perturbed
the Mn-C\textsubscript{6}N\textsubscript{6} system with an applied perpendicular electric field ranging from 0.0 to 10
V/nm. Under each applied electric field strength, the Mn-C\textsubscript{6}N\textsubscript{6} system was
fully-optimized. The Mn-C\textsubscript{6}N\textsubscript{6} system slightly buckles after optimization. The buckling
becomes more prominent at the maximum value of 10 V/nm (bottom figure in Fig.~\ref{fig:fig1}(b)). Related to the buckling, the
binding energy calculated at each applied electric field strength shows an increasing pattern (see Fig.~\ref{fig:fig2}(b)). This
indicates that the larger the applied electric field is, the stronger the binding of Mn atom in the porous site
becomes. It is expected that the semiconducting property of Mn-C\textsubscript{6}N\textsubscript{6} system would change
under the application of electric fields, as reported for some related 2D materials [53]. However, despite the small
wrinkles, the electronic and magnetic properties of Mn-C\textsubscript{6}N\textsubscript{6} are retained up to a peak
value of 10 V/nm. To show that the semiconducting property remains unchanged, we have plotted the TDOS of
Mn-C\textsubscript{6}N\textsubscript{6} at a maximum of 10 V/nm (see Fig.~\ref{fig:fig2} (d)). This can be obviously seen by
comparing Fig.~\ref{fig:fig2}(d) and Fig.~\ref{fig4} (c). Specifically, the semiconducting property remains unchanged.}

{ 
The appearance of wrinkles in the Mn- C\textsubscript{6}N\textsubscript{6} under applied electric field is due to
distortion caused by  $d_{\mathit{xy}}$,  $d_{x^2-y^2}$ orbitals of Mn in trying to lower the system's energy. As a
result, the hexagonal structure is distorted. We refer to the projected density of states plots (see Figs.~\ref{fig4} (d)-(f))
where we show the orbital hybridizations between  $p_x,p_y$ orbitals of N and  $d_{\mathit{xy}}$,  $d_{x^2-y^2}$ of Mn.
This clearly indicates that both the orbitals are on the same plane. Due to repulsion between the orbitals, the system
energy becomes lower. These conditions of having preserved electronic and magnetic properties under applied electric
field of Mn-C\textsubscript{6}N\textsubscript{6} sheet could be used in future nano-devices.}

\subsection{Adsorption of atoms on Mn C\textbf{\textsubscript{6}}{N}{\textsubscript{6}}{sheet}}

{ 
We first relaxed the Mn-C\textsubscript{6}N\textsubscript{6} with each type of atoms (C, N, O, H) adsorbed at a height
of 2 \AA\textit{ } above Mn atom in the sheet. We do not impose any geometry constraint during structural relaxation. After
relaxation, the Mn-C\textsubscript{6}N\textsubscript{6} sheets remain planar with no sign of wrinkles in the structures
(see Figs. 5 (i)-(iv). The optimized structural properties and corresponding adsorption energies are listed in Table~\ref{table3}.
We find that the calculated optimized bond length between the embedded Mn atom and the adsorbed atoms, 
$d_{\mathrm{Mn}-X}$ ( $X$\textcolor{black}{ stands for atoms or molecules)}, varies only slightly in the range
1.70-1.92 \AA\textit{ }. This range of value confirms the expected covalent bond length between the Mn and the adatoms (adsorbed
atoms). To quantify the extent of chemisorption of the adsorbates on Mn-C\textsubscript{6}N\textsubscript{6} sheet, we
compute their adsorption energies defined in the Eq. (5) below,}

\begin{equation}
E_{\mathrm{ads}}=(E_{\mathrm{Mn}-\mathrm{C}_6\mathrm{N}_6} + E_X)-E_T
\end{equation}
where \textit{E}\textit{\textsubscript{T}} is the total energy of Mn-C\textsubscript{6}N\textsubscript{6} with
adsorbates, $E_{\mathrm{Mn}-\mathrm{C}_6\mathrm{N}_6}$ represents the energy of Mn-C\textsubscript{6}N\textsubscript{6} system, and \textit{E}\textsubscript{X} is the total energy of an isolated atom
or molecules. Positive adsorption energy shows a stable structure. As illustrated in Table~\ref{table3}, N atom has the lowest
adsorption energy and the weakest covalent bond length. This can be related to the nature of interaction between the Mn
and N atom, in which nitrogen is the most electronegative among the adatoms. In general, the adsorption energies and
their corresponding bond lengths show that the structures are stable and all atoms are chemically bonded to the
embedded Mn atom in the C\textsubscript{6}N\textsubscript{6} sheet.

{ 
The total magnetic moment per unit cell and per atom are also shown in Table~\ref{table3}. Unlike the
Mn-C\textsubscript{6}N\textsubscript{6} system, the magnetic moment per unit cell are found to be reduced when atoms
are adsorbed on the Mn-C\textsubscript{6}N\textsubscript{6} sheet. The order of magnetic moment of
Mn-C\textsubscript{6}N\textsubscript{6} with adatoms is C{\textgreater}N{\textgreater}O{\textgreater}H. As can be seen
in Table~\ref{table3}, magnetism is mainly contributed by the Mn atom, and the adatoms couple antiferromagnetically with the
embedded Mn atom. To clarify the reason for reduction in total magnetic moment per unit cell, we have evaluated the
charge transfer into the C\textsubscript{6}N\textsubscript{6} sheet and adatoms in the structure. It can be clearly
seen that the reduction is a consequence of electron injection into the C\textsubscript{6}N\textsubscript{6} sheet from
the adatoms and Mn. Such an electron redistribution between the orbitals of Mn and the surrounding atoms induces strong
interaction. The interaction reduces the number of spin-polarized electrons of Mn atom. Consequently, these low-spin
configurations lead to the reduction of the system's total magnetic moment. The TDOS of
Mn-C\textsubscript{6}N\textsubscript{6} with adsorbed atoms are depicted in Fig.~\ref{fig5}. In spite of hybridization between
the adatoms and the embedded Mn atom, the electronic structures of the majority of the systems remain semiconducting
with their respective band gaps listed in Table~\ref{table3}. The TDOS for Mn-C\textsubscript{6}N\textsubscript{6} with adsorbed O
atom shows half metallic electronic character.}

\subsection{Adsorption of molecules on Mn-C{\textsubscript{6}}{N}{\textsubscript{6}} {sheet}}

{ 
Six molecules (CH\textsubscript{4}, N\textsubscript{2}, O\textsubscript{2}, H\textsubscript{2}, CO, CO\textsubscript{2})
adsorbed at a height of 2  $\text{{\AA\textit{ }}}$ above the porous sites in the Mn-C\textsubscript{6}N\textsubscript{6} sheet
are considered. We do not impose any geometry constraint during structural relaxation. As depicted in Fig.~\ref{fig6} (i) --
(vi), the Mn-C\textsubscript{6}N\textsubscript{6} sheet remains planar with no sign of buckling in the structure after
relaxation. \ The optimized bond length between the embedded Mn atom and the nearest atom in the adsorbed molecules 
$d_{\mathrm{Mn}-X}$\textit{ }are in the range 2.1-2.6  $\text{{\AA\textit{ }}}$ and the adsorption energies are in the range of
0.11 - \~{} 0.76 eV. The optimized bond lengths for isolated molecules in all cases, except oxygen, are maintained
after adsorption. Based on the previous report [41], the interaction between embedded 3d TM and O\textsubscript{2} is
usually strong; we have confirmed this assertion for the embedded Mn atom with adsorbed O\textsubscript{2}. We note
that Mn-C\textsubscript{6}N\textsubscript{6} with adsorbed O\textsubscript{2} has the highest adsorption energy and
lowest optimized bond length,  $d_{\mathrm{Mn}-\mathrm{O}_2}$, due to strong chemisorption. The bond length of the isolated
O\textsubscript{2} is slightly lower than the bond length of the adsorbed O\textsubscript{2}. This finding is
consistent with the previous work [41]. }

{ 
Interestingly,  $d_{\mathrm{Mn}-\mathrm{CO}}$ is the largest among all  $d_{\mathrm{Mn}-X}$, showing that CO is the
weakest of all molecules that bind to the surface., Correspondingly, its adsorption energy is also low. The weak
chemisorption energy for CO confirms the previously reported assertion that O\textsubscript{2} would preferentially
cover the Mn-C\textsubscript{6}N\textsubscript{6} as compared to CO when the two gases are simultaneously released at
constant pressure on the surface [41]. Their conclusion was reached based on the calculation that assumes
O\textsubscript{2} has larger adsorption energy as compared to that of CO molecule. Here we would like to contribute to
the understanding of adsorption of O\textsubscript{2} and CO based on our previous computations of binding energies of
Mn-C\textsubscript{6}N\textsubscript{6} systems under bi-axial tensile strain and applied perpendicular electric field.
The binding energy is found to reduce under symmetric deformation induced by bi-axial tensile strain, and increase as
larger electric field strength is applied. Therefore, these external environments can be used to modulate the
adsorption energy. For example, the adsorption energy for O\textsubscript{2} shows decreasing pattern from 0.789 eV @
0.5\% to 0.658 @ 1\% as larger symmetric deformations (biaxial tensile strain) are applied. Correspondingly, adsorption
energy of CO on Mn-C\textsubscript{6}N\textsubscript{6} becomes larger from 0.189 eV @ 1 V/nm to 0.243 @ 2 V/nm\textbf{
}as larger electric field strengths are applied. We hope that the two methods of modulating the adsorption energy could
be useful in the single crystal field (SAC) application.}

{ 
The adsorption energies for CH\textsubscript{4}, N\textsubscript{2}, H\textsubscript{2} molecules on
Mn-C\textsubscript{6}N\textsubscript{6} are in agreement with recent adsorption energies reported for the same
molecules adsorbed on pure C\textsubscript{6}N\textsubscript{6} [54]. However, due to interactions between the embedded
Mn atom and the molecules, the corresponding  $d_{\mathrm{Mn}-X}$ are found to be lower. Our calculated adsorption
energy for CO\textsubscript{2} is within the value suggested by Deng \textit{et al.} [55], and the molecule is slightly
physisorbed on Mn-C\textsubscript{6}N\textsubscript{6} sheet.}

{ 
The total magnetic moments per unit cell for molecular adsorption on Mn-C\textsubscript{6}N\textsubscript{6} are
comparable to Mn-C\textsubscript{6}N\textsubscript{6} except for the O\textsubscript{2} molecule. As for previous
cases, this observation can be explained in terms of electron redistribution between the orbitals in Mn atom and the
surrounding atoms (O adatoms included) induced by strong interaction among themselves. This causes the number of
spin-polarized electrons of Mn atom to become reduced. The consequent low-spin configurations resulted from reduction
in number of unpaired electron renders the total magnetic moment to reduce. We have also calculated the charge transfer
into the C\textsubscript{6}N\textsubscript{6} sheet and adatoms in the structure. It is evident that the reduced
unpaired electrons are resulted from the electron injection into the C\textsubscript{6}N\textsubscript{6} sheet and the
adatoms from Mn atom. The TDOS of Mn-C\textsubscript{6}N\textsubscript{6} with adsorbed molecules are shown in Fig.~\ref{fig6}
(i) -- (vi). Except Mn-C\textsubscript{6}N\textsubscript{6} with adsorbed O\textsubscript{2} molecule showing half
metallic character, all the remaining systems have their semiconducting electronic structure maintained. The band gaps
are listed in Table~\ref{table3}.}

\section{Conclusions}

In this work, mechanical, structural, electronic and magnetic properties of C\textsubscript{6}N\textsubscript{6} with
embedded Mn atom is studied systematically under the influence of external environment using density functional theory
calculation. The Mn-C\textsubscript{6}N\textsubscript{6} is found to be structurally and mechanically stable. Our
results show that the binding energy of Mn embedded in C\textsubscript{6}N\textsubscript{6} decrease as larger tensile
strain is applied. The decrease in binding energy can be related to the structural distortion of the
C\textsubscript{6}N\textsubscript{6} sheet. We find that the semiconducting property of the pure sheet is preserved
after the embedment of Mn atom in porous site. We have shown a linear relation in the increment of band gap and applied
bi-axial tensile strain imposed on the Mn-C\textsubscript{6}N\textsubscript{6} systems. Magnetic moment of the embedded
Mn atom is preserved in all cases. Additionally, Mn-C\textsubscript{6}N\textsubscript{6} sheet slightly buckles after
structural relaxation under the influence of applied perpendicular electric field. Despite the obvious wrinkles in the
Mn-C\textsubscript{6}N\textsubscript{6} structure, the electronic and magnetic properties of the
Mn-C\textsubscript{6}N\textsubscript{6} systems are maintained up to a peak value of 10 V/nm in electric field
strength. We also investigate the geometries, electronic and magnetic properties of
Mn-C\textsubscript{6}N\textsubscript{6} with adsorbed atoms and molecules. Except for
Mn-C\textsubscript{6}N\textsubscript{6} with adsorbed oxygen atom and molecule which show half metallic character, the
remaining systems maintained the semiconducting property of the Mn-C\textsubscript{6}N\textsubscript{6} sheet. The
total magnetic moment per unit cell in most of the systems is found to reduce as compared to that for
Mn-C\textsubscript{6}N\textsubscript{6} sheet. The reduction in magnetic moment can be related to the strong
interactions among the Mn atom and the surrounding atoms which lead to the formation of low-spin configurations. In
general, these results indicate that Mn-C\textsubscript{6}N\textsubscript{6} systems with and without adsorbed atoms
and molecules can serve as potential candidates for future spintronics and catalysis applications.

\section*{Acknowledgments}
T. L. Yoon wishes to acknowledge the support of Universiti Sains Malaysia RU grant (No. 1001/PFIZIK/811240). Figures
showing atomic model and 2D charge-density difference plots are generated using the XCRYSDEN program Ref. [56]. We
gladfully acknowledge Dr. Chan Huah Yong from the School of Computer Science, USM, for providing us computing resources
to carry out part of the calculations done in this paper.

\bigskip
\noindent { \textbf{References}}\\

{ 
[1]\ \ M.N. Baibich, J.M. Broto, A. Fert, F.N. Van Dau, F. Petroff, P. Etienne, G. Creuzet, A. Friederich, J. Chazelas,
Phys. Rev. Lett., 61 (1988) 2472.}

{ 
[2]\ \ G. Binasch, P. Grunberg, F. Saurenbach, W. Zinn, Phys. Rev. B., 39 (1989) 4828.}

{ 
[3]\ \ Rollmann, G., Sahoo, S., Hucht, A., \& Entel, P. Phys. Rev. B., 78 (2008) 134404.}

{ 
[4]\ \ Zeleny, M., \v{S}ob, M., \& Hafner, J. Phys. Rev. B., 80 (2009) 144414.}

{ 
[5]\ \ D. Duffy, J. Blackman, Phys. Rev. B., 58 (1998) 7443.}

{ 
[6]\ \ Y.Z. Abdullahi, M. Rahman, S. Abubakar, H. Setiyanto, Quantum Matter, 4 (2015) 430.}

{ 
[7]\ \ M. Rahman, Y.Z. Abdullahi, A. Shuaibu, S. Abubakar, H. Zainuddin, R. Muhida, H. Setiyanto, J. Comput. Theor.
Nanosci., 12 (2015) 1995.}

{ 
[8]\ \ V. Stepanyuk, W. Hergert, K. Wildberger, R. Zeller, P. Dederichs, Phys. Rev. B., 53 (1996) 2121.}

{ 
[9]\ \ B. Lazarovits, L. Szunyogh, P. Weinberger, Phys. Rev. B., 65 (2002) 104441.}

{ 
[10]\ \ A. AlZahrani, \textstyleStrong{\textmd{Physica B Condens Matter.,}} 407 (2012) 992.}

{ 
[11]\ \ Y.Z. Abdullahi, M.M. Rahman, A. Shuaibu, S. Abubakar, H. Zainuddin, R. Muhida, H. Setiyanto,
\textstyleStrong{\textmd{Physica B Condens Matter.,}} 447 (2014) 65.}

{ 
[12]\ \ Z. Zhang, Z. Geng, D. Cai, T. Pan, Y. Chen, L. Dong, T. Zhou, Physica E., 65 (2015) 24.}

{ 
[13]\ \ Y. Yagi, T.M. Briere, M.H. Sluiter, V. Kumar, A.A. Farajian, Y. Kawazoe, Phys. Rev. B., \ 69 (2004) 075414.}

{ 
[14]\ \ E. Kroke, M. Schwarz, E. Horath-Bordon, P. Kroll, B. Noll, A.D. Norman, New J. Chem., 26 (2002) 508.}

{ 
[15]\ \ Y. Z. Abdullahi, T. L. Yoon, M. M. Halim, M. R. Hashim, T. L. Lim, Solid State Commun., 248 (2016) 144.}

{ 
[16]\ \ A. Wang, X. Zhang, M. Zhao, Nanoscale 6 (2014) 11157.}

{ 
[17]\ \ D.M. Teter, R.J. Hemley, Science 271 (1996) 53.}

{ 
[18]\ \ Y.Z. Abdullahi, T.L. Yoon, M.M. Halim, M.R. Hashim, M.Z.M. Jafri, L.T. Leng, \textstylest{Curr. Appl. Phys}., 16
(2016) 809.}

{ 
[19]\ \ S.M. Aspera, H. Kasai, H. Kawai, Surf. Sci., 606 (2012) 892.}

{ 
[20]\ \ S. Zhang, R. Chi, C. Li, Y. Jia, Phys. Lett. A., 380 (2016) 1373.}

{ 
[21]\ \ D. Ghosh, G. Periyasamy, B. Pandey, S.K. Pati, J. Mater. Chem. C., 2 (2014) 7943.}

{ 
[22]\ \ J. Tian, Q. Liu, A.M. Asiri, A.O. Al-Youbi, X. Sun, Anal. Chem., 85 (2013) 5595.}

{ 
[23]\ \ P. Niu, L. Zhang, G. Liu, H.M. Cheng, Adv. Funct. Mater., 22 (2012) 4763.}

{ 
[24]\ \ N. Fechler, G.A. Tiruye, R. Marcilla, M. Antonietti, RSC Adv., 4 (2014) 26981.}

{ 
[25]\ \ K. Sridharan, P. Sreekanth, T.J. Park, R. Philip, J. Phys. Chem. C., 119 (2015) 16314.}

{ 
[26]\ \ B. Meng, W.-z. Xiao, L.-l. Wang, L. Yue, S. Zhang, H.-y. Zhang, Phys. Chem. Chem. Phys., 17 (2015) 22136.}

{ 
[27]\ \ I. Choudhuri, P. Garg, B. Pathak, J. Mater. Chem. C., 4 (2016) 8253.}

{ 
[28]\ \ J. Du, C. Xia, W. Xiong, X. Zhao, T. Wang, Y. Jia, Phys. Chem. Chem. Phys., 18 (2016) 22678.}

{ 
[29]\ \ S. Sun, G. Zhang, N. Gauquelin, N. Chen, J. Zhou, S. Yang, W. Chen, X. Meng, D. Geng, M.N. Banis, Sci. Rep., 3
(2013).}

{ 
[30]\ \ Y. Ji, H. Dong, H. Lin, L. Zhang, T. Hou, Y. Li, RSC Adv., 6 (2016) 52377.}

{ 
[31]\ \ Z. Ma, X. Zhao, Q. Tang, Z. Zhou, Int. J. Hydrogen Energy., 39 (2014) 5037.}

{ 
[32]\ \ X. Liu, T. Duan, Y. Sui, C. Meng, Y. Han, RSC Adv., 4 (2014) 38750.}

{ 
[33]\ \ S. Sinthika, E.M. Kumar, R. Thapa, J. Mater. Chem. A., 2 (2014) 12812.}

{ 
[34]\ \ S. Stolbov, M. Alcantara Ortigoza, J. Chem. Phys., 142 (2015) 154703.}

{ 
[35]\ \ G. Vile, D. Albani, M. Nachtegaal, Z. Chen, D. Dontsova, M. Antonietti, N. Lopez, J. Perez{}-Ramirez, Angew.
Chem. Int. Ed., 54 (2015) 11265.}

{ 
[36]\ \ P. Wu, P. Du, H. Zhang, C. Cai, Phys. Chem. Chem. Phys., 17 (2015) 1441.}

{ 
[37]\ \ F. Ersan, O. Arslanalp, G. Gokoglu, E. Akturk, Appl. Surf. Sci., 371 (2016) 314.}

{ 
[38]\ \ Y. Li, Z. Zhou, G. Yu, W. Chen, Z. Chen, J. Phys. Chem. C., 114 (2010) 6250.}

{ 
[39]\ \ Y. Tang, Z. Yang, X. Dai, Phys. Chem. Chem. Phys., \ 14 (2012) 16566.}

{ 
[40]\ \ H. Yan, H. Cheng, H. Yi, Y. Lin, T. Yao, C. Wang, J. Li, S. Wei, J. Lu, J. Am. Chem. Soc., 137 (2015) 10484.}

{ 
[41]\ \ D. Ma, Q. Wang, X. Yan, X. Zhang, C. He, D. Zhou, Y. Tang, Z. Lu, Z. Yang, Carbon 105 (2016) 463.}

{ 
[42]\ \ P. Hohenberg, W. Kohn, Phys Rev., 136 (1964) B864.}

{ 
[43]\ \ P. Giannozzi, S. Baroni, N. Bonini, M. Calandra, R. Car, C. Cavazzoni, D. Ceresoli, G.L. Chiarotti, M.
Cococcioni, I. Dabo, J. Phys. Condens. Matter., 21 (2009) 395502. }

{ 
[44]\ \ J.P. Perdew, K. Burke, M. Ernzerhof, Phys. Rev. Lett., 77 (1996) 3865.}

{ 
[45]\ \ M. Cococcioni, S. De Gironcoli, Phys. Rev. B., 71 (2005) 035105.}

{ 
[46]\ \ D. Vanderbilt, Phys. Rev. B., 41 (1990) 7892. }

{ 
[47]\ \ D. Vanderbilt, Phys. Rev. B., 41 (1990) 7892.}

{ 
[48]\ \ H.J. Monkhorst, J.D. Pack, Phys. Rev. B., 13 (1976) 5188.}

{ 
[49]\ \ M. Asadpour, S. Malakpour, M. Faghihnasiri, B. Taghipour, Solid State Commun., 212, 46 (2015).}

{ 
[50]\ \ E. Cadelano, P.L. Palla, S. Giordano, L. Colombo, Phys. Rev. B., 82 (2010) 235414.}

{ 
[51]\ \ K. Srinivasu, B. Modak, S.K. Ghosh, Phys. Chem. Chem. Phys., 18 (2016) 26466.}

{ 
[52]\ \ P.O. Lowdin, J. Chem. Phys., 18 (1950) 365.}

{ 
[53]\ \ B. Huang, H. Xiang, J. Yu, S.-H. Wei, Phys. Rev. Lett., 108 (2012) 206802.}

{ 
[54]\ \ Ma, Z., Zhao, X., Tang, Q., \& Zhou, Z, Int. J. Hydrogen Energy., 39 (2014) 5037-5042.}

{ 
[55]\ \ Q. Deng, L. Zhao, X. Gao, M. Zhang, Y. Luo, Y. Zhao, Small 9 (2013) 3506.}

{ 
[56]\ \ A. Kokalj, Computer. Mater. Sci.,\textit{ }\ 28 (2003) 155.}


\begin{table}[!p]
\caption{Optimized lattice parameters and total strain energy of Mn-C\textsubscript{6}N\textsubscript{6} system for
mechanical properties computation.}

\begin{center}
\tablefirsthead{}
\tablehead{}
\tabletail{}
\tablelasttail{}
\begin{supertabular}{m{0.6136598in}m{0.8205598in}m{1.0344598in}m{0.9684598in}m{0.8677598in}}
\hline
{\centering  Strain\par}

\centering{  (\%)} &
{\centering  Area\par}

\centering{  (\AA\textit{ }\textsuperscript{2})} &
{\centering  Total energy\par}

{\centering  Biaxial\par}

\centering{  (Ry)} &
{\centering  Total energy\par}

{\centering  Uniaxial\par}

\centering{  (Ry)} &
{\centering  Lattice parameter\par}

{\centering  Biaxial\par}

\centering\arraybslash{  (\AA\textit{ })}\\\hline
\centering{  {}-0.02} &
\centering{  167.61} &
\centering{  {}-959.07561 \ } &
{  {}-959.10937} &
\centering\arraybslash{  13.12/24.10}\\
\centering{  {}-0.015} &
\centering{  169.32} &
\centering{  {}-959.10183} &
{  {}-959.12084} &
\centering\arraybslash{  13.98/24.22 \ \ }\\
\centering{  {}-0.01} &
\centering{  171.05} &
\centering{  {}-959.12023} &
{  {}-959.12875} &
\centering\arraybslash{  14.05/24.34 \ \ }\\
\centering{  {}-0.005} &
\centering{  172.78} &
\centering{  {}-959.13109} &
{  {}-959.13287} &
\centering\arraybslash{  14.13/24.47}\\
\centering{  0} &
\centering{  174.52} &
\centering{  {}-959.13465} &
{  {}-959.13465} &
\centering\arraybslash{  14.20/24.59}\\
\centering{  0.005} &
\centering{  176.27} &
{  \ \ \ \ {}-959.13151} &
{  {}-959.13342} &
\centering\arraybslash{  14.27/24.71}\\
\centering{  0.01} &
\centering{  178.03} &
\centering{  {}-959.12150 } &
{  {}-959.12937} &
\centering\arraybslash{  14.34/24.83}\\
\centering{  0.015} &
{  \ \ \ \ \ 179.79 } &
\centering{  {}-959.10532} &
{  {}-959.12240} &
\centering\arraybslash{  14.41/24.96}\\
\centering{  0.02} &
\centering{  181.57} &
\centering{  {}-959.08281} &
{  {}-959.11258} &
\centering\arraybslash{  14.48/25.08}\\\hline
\end{supertabular}
\end{center}
\label{table1}
\end{table}

\begin{table}[!p]
\caption{Geometry and electronic structure for strained/unstrained Mn-C\textsubscript{6}N\textsubscript{6}. The
binding energies \textit{E}\textit{\textsubscript{b}}, the average bond length between Mn atom and
N\textsubscript{edge} atoms, average bond length connecting the \textit{s}{}-triazine, and Mn height are respectively
named as \textit{d}\textsubscript{Fe-N}, \textit{d}\textsubscript{1}/\textit{d}\textsubscript{2}, and \textit{h }(is
the difference in the \textit{z}{}-coordinate of the Mn atom and the average of the \textit{z}{}-coordinate of all the
C and N atoms in the C\textsubscript{6}N\textsubscript{6} sheet). Charge transfer, magnetic moment per unit cell and
per Mn atom,\textcolor{black}{ electronic character of the }Mn-C\textsubscript{6}N\textsubscript{6}\textcolor{black}{
system are denoted by }\textit{\textcolor{black}{Q}}\textcolor{black}{,
M}\textcolor{black}{\textsubscript{cell}}\textcolor{black}{, M}\textcolor{black}{\textsubscript{Fe}}\textcolor{black}{,
EC respectively. All the systems are semiconducting, }Values in the EC column denotes that of the semiconducting band
gap\textcolor{black}.}

\begin{center}
\tablefirsthead{}
\tablehead{}
\tabletail{}
\tablelasttail{}
\begin{supertabular}{m{0.5545598in}m{0.46015987in}m{0.44415984in}m{0.59625983in}m{0.51365983in}m{0.7851598in}m{0.37685984in}m{0.38515985in}m{0.44625983in}}
\hline
\centering{  Strain} &
{\centering  \textit{E}\textsubscript{b}\par}

\centering{  (eV)} &
{\centering  \textit{d}\textsubscript{Mn-N}\par}

\centering{  (\AA\textit{ })} &
{\centering  \textit{d}\par}

\centering{  (\AA\textit{ })} &
{\centering  \textit{H}\par}

\centering{  (\AA\textit{ })} &
{\centering  Q\par}

\centering{  (electrons)} &
{\centering  M\textsubscript{Mn}\par}

\centering{  (\textit{{\textmu}}\textsubscript{B})} &
{\centering  M\textsubscript{cell}\par}

\centering{  (\textit{{\textmu}}\textsubscript{B})} &
\centering\arraybslash{  EC}\\\hline
\centering{  0\%} &
\centering{  4.14} &
\centering{  2.70} &
\centering{  1.48} &
\centering{  {}-0.011} &
\centering{  0.75} &
\centering{  5.00} &
\centering{  5.02} &
\centering\arraybslash{  0.630}\\
\centering{  1\%} &
\centering{  3.97} &
\centering{  2.75} &
\centering{  1.50} &
\centering{  {}-0.007} &
\centering{  0.75} &
\centering{  5.01} &
\centering{  5.02} &
\centering\arraybslash{  0.660}\\
\centering{  2\%} &
\centering{  3.44} &
\centering{  2.79} &
\centering{  1.52} &
\centering{  0.003} &
\centering{  0.76} &
\centering{  5.02} &
\centering{  5.02} &
\centering\arraybslash{  0.675}\\
\centering{  3\%} &
\centering{  2.60} &
\centering{  2.84} &
\centering{  1.55} &
\centering{  0.010} &
\centering{  0.76} &
\centering{  5.00} &
\centering{  5.03} &
\centering\arraybslash{  0.764}\\
\centering{  4\%} &
\centering{  1.48} &
\centering{  2.89} &
\centering{  1.58} &
\centering{  0.014} &
\centering{  0.77} &
\centering{  5.04} &
\centering{  5.04} &
\centering\arraybslash{  0.807}\\
\centering{  5\%} &
\centering{  0.10} &
\centering{  2.93} &
\centering{  1.60} &
\centering{  0.000} &
\centering{  0.77} &
\centering{  5.06} &
\centering{  5.05} &
\centering\arraybslash{  0.802}\\
\centering{  6\%} &
\centering{  {}-1.49} &
\centering{  2.99} &
\centering{  1.63} &
\centering{  0.003} &
\centering{  0.77} &
\centering{  5.09} &
\centering{  5.07} &
\centering\arraybslash{  0.952}\\\hline
\end{supertabular}
\end{center}
\label{table2}
\end{table}

\begin{table}[!p]
\caption{Geometry and electronic structure for Mn-C\textsubscript{6}N\textsubscript{6} with adsorbed
atoms/molecules.  $X$ refers to an adsorbate species.  $E_{\mathit{ads}}$ refers to adsorption energy, 
$d_{\mathrm{Mn}-X}$ refers to averaged bond length between Mn atom and the adsorbates.  $d_{\mathit{avg}-X}$ refers to
bond length of molecules. The values without parenthesis are that for absorbed molecules while that in parenthesis are
for isolated molecules.  $Q$ refers to net charge transfer among the adsorbates and the
Mn-C\textsubscript{6}N\textsubscript{6 }sheet. The values without parenthesis are charge transfer from Mn atom into the
sheet or adsorbates. These values are all positive as electron from Mn atom always get transferred into the
surrounding. The  $Q$ values in parenthesis are charge transfer into the Mn-C\textsubscript{6}N\textsubscript{6 }sheet
from adsorbates. Positive values mean electron is transferred into the surroundings
(Mn-C\textsubscript{6}N\textsubscript{6 }sheet) from adsorbates, and vice versa.  $M_{\mathit{cell}}$ refers to
magnetic moment per unit cell.  $M_{\mathit{atom}}$ refers to magnetic moment of Mn atom or adsorbates. The values
without parenthesis are that for Mn atom, while that in parenthesis are that for the adsorbates. EC refers to the
electronic character of the Mn-C\textsubscript{6}N\textsubscript{6} with adsorbates. In the present case, EC can be
either half metallic (HM) or semiconducting (SC). The values in the SC case are that of its corresponding band gap. }

\begin{center}
\tablefirsthead{}
\tablehead{}
\tabletail{}
\tablelasttail{}
\begin{supertabular}{m{0.5615598in}m{0.41845986in}m{0.5448598in}m{0.6080598in}m{1.4240599in}m{0.38025984in}m{0.76435983in}m{0.5566598in}}
\hline
\centering{  System} &
{\centering  \textit{E}\textsubscript{ads}\par}

\centering{  (eV)} &
{\centering  \textit{d}\textsubscript{Mn-X}\par}

\centering{  (\AA\textit{ })} &
{\centering  \textit{d}\textit{\textsubscript{avg-X}}\par}

\centering{  (\AA\textit{ })} &
{\centering  \textit{Q}\par}

\centering{  (electrons)} &
{\centering  M\textsubscript{cell}\par}

\centering{  (\textit{{\textmu}}\textsubscript{B})} &
{\centering  M\textsubscript{atom}\par}

\centering{  (\textit{{\textmu}}\textsubscript{B})} &
\centering\arraybslash{  EC}\\\hline
\centering{  C} &
\centering{  1.98} &
\centering{  1.91} &
\centering{  {}-} &
{\centering  0.40\par}

\centering{  (0.13)} &
\centering{  2.25} &
{\centering  4.56\par}

\centering{  (-2.25)} &
\centering\arraybslash{  SC (0.843)}\\
\centering{  N} &
\centering{  1.37} &
\centering{  2.54} &
\centering{  {}-} &
{\centering  0.58\par}

\centering{  (0.18)} &
\centering{  2.68} &
{\centering  4.76\par}

\centering{  (-2.11)} &
{\centering  SC\par}

\centering\arraybslash{  (0.440)}\\
\centering{  O} &
\centering{  2.53} &
\centering{  1.89} &
\centering{  {}-} &
{\centering  0.65\par}

\centering{  (0.41)} &
\centering{  3.89} &
{\centering  4.83\par}

\centering{  (-0.98)} &
\centering\arraybslash{  HM}\\
\centering{  H} &
\centering{  1.51} &
\centering{  1.70} &
\centering{  {}-} &
{\centering  0.48\par}

\centering{  (0.16)} &
\centering{  4.78} &
{\centering  4.92\par}

\centering{  (-0.20)} &
{\centering  SC\par}

\centering\arraybslash{  (0.900)}\\
\centering{  CO} &
\centering{  0.16} &
\centering{  2.66} &
{\centering  1.14\par}

\centering{  (1.15)} &
{\centering  0.67\par}

\centering{  (C: -0.30, O: 0.07)} &
\centering{  5.00} &
{\centering  4.98\par}

\centering{  (0.00)} &
{\centering  SC\par}

\centering\arraybslash{  (0.793)}\\
\centering{  CO\textsubscript{2}} &
\centering{  0.35} &
\centering{  2.40} &
{\centering  1.18\par}

\centering{  (1.18)} &
{\centering  0.69\par}

\centering{  (C: -0.79, O: -0.43)} &
\centering{  5.00} &
{\centering  4.97\par}

\centering{  (0.00)} &
{\centering  SC\par}

\centering\arraybslash{  (0.822)}\\
\centering{  O\textsubscript{2}} &
\centering{  0.76} &
\centering{  2.12} &
{\centering  1.23\par}

\centering{  (1.32)} &
{\centering  0.66\par}

\centering{  (0.20)} &
\centering{  3.48} &
{\centering  4.90\par}

\centering{  (-1.43)} &
\centering\arraybslash{  HM}\\
\centering{  N\textsubscript{2}} &
\centering{  0.11} &
\centering{  2.54} &
{\centering  1.12\par}

\centering{  (1.11)} &
{\centering  0.62\par}

\centering{  (-0.28)} &
\centering{  4.99} &
{\centering  4.98\par}

\centering{  (0.00)} &
{\centering  SC\par}

\centering\arraybslash{  (0.810)}\\
\centering{  H\textsubscript{2}} &
\centering{  0.11} &
\centering{  2.38} &
{\centering  0.75\par}

\centering{  (0.76)} &
{\centering  0.69\par}

\centering{  (-0.16)} &
\centering{  4.99} &
{\centering  4.98\par}

\centering{  (0.00)} &
{\centering  SC\par}

\centering\arraybslash{  (0.652)}\\
\centering{  CH\textsubscript{4}} &
\centering{  0.13} &
\centering{  2.54} &
{\centering  1.10\par}

\centering{  (1.10)} &
{\centering  0.72\par}

\centering{  (C: 0.67, H: -0.88)} &
\centering{  5.00} &
{\centering  4.99\par}

\centering{  (0.00)} &
{\centering  SC\par}

\centering\arraybslash{  (0.708)}\\\hline
\end{supertabular}
\end{center}
\label{table3}
\end{table}

%

\begin{figure}*[!p]
\begin{center}
(a) \includegraphics[width=2.7in,height=2.7in]{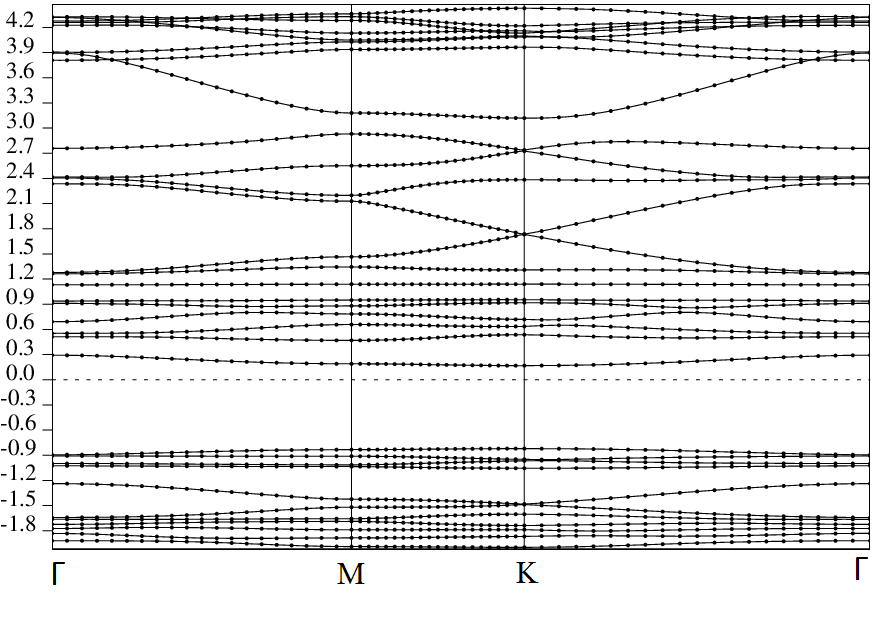}
(b) \includegraphics[width=2.7in,height=2.7in]{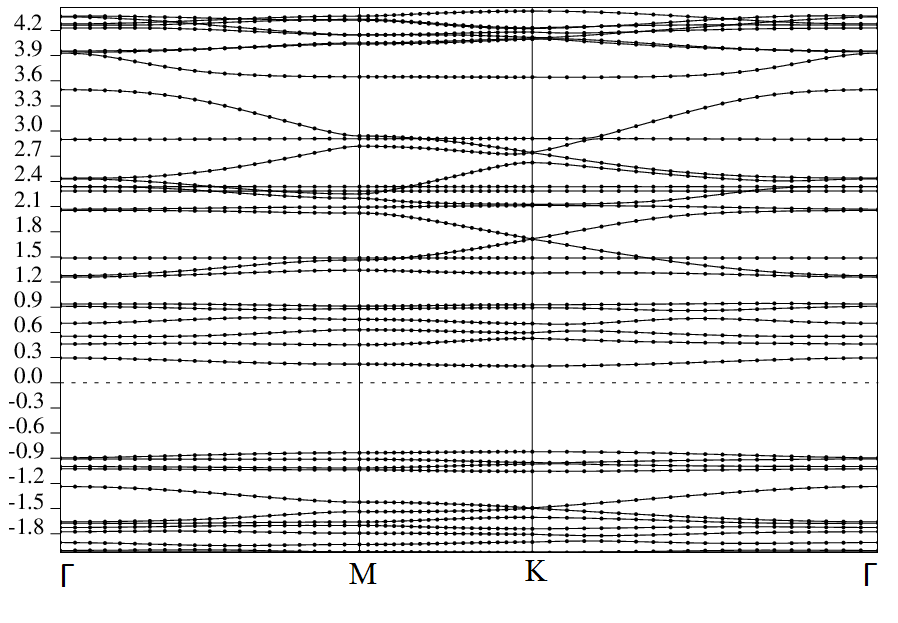}
\end{center}
\begin{center}
\includegraphics[width=6.3126in,height=2.7in]{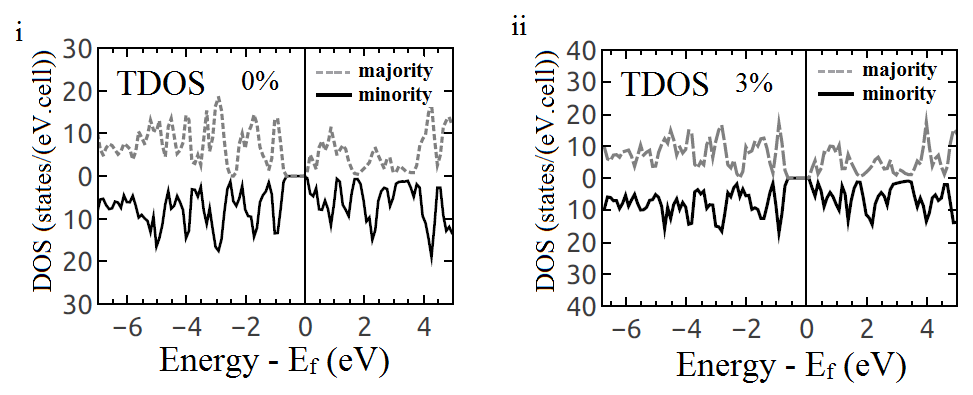}
\end{center}
(c)
\begin{center}
{\centering \includegraphics[width=6.3126in,height=2.7in]{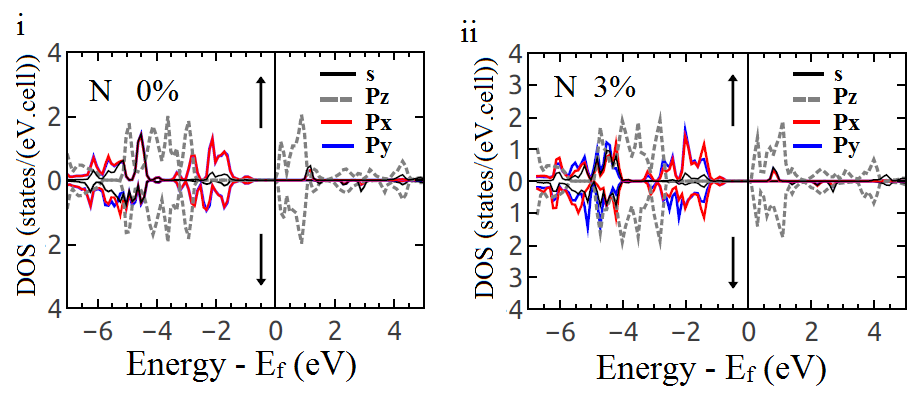} }
\end{center}
(d)
\end{figure}*

\begin{figure}[!p]
\begin{center}
{\centering \includegraphics[width=6.3126in,height=2.7in]{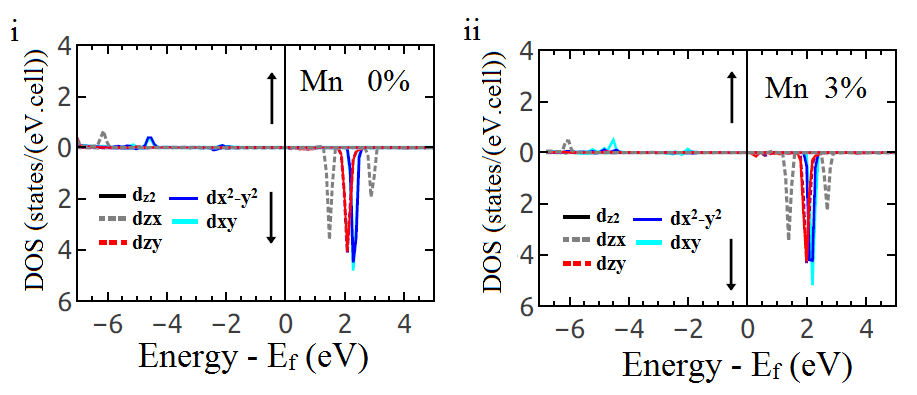}}
\end{center}
(e)
\begin{center}
{\centering \includegraphics[width=6.3126in,height=2.7in]{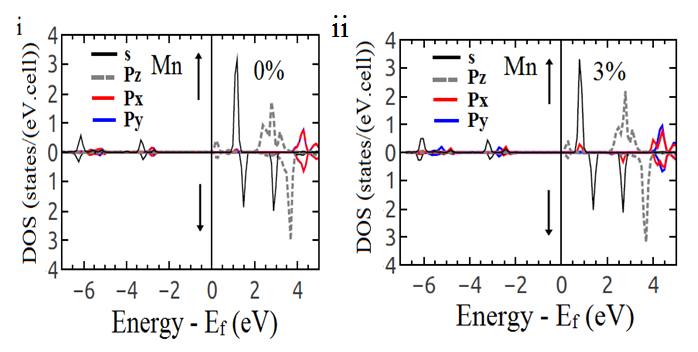}}
\end{center}
(f)
\caption{Spin-polarized electronic band structure (a) majority (b) minority spin states for unstrained (\textit{s} = 0) Mn-C\textsubscript{6}N\textsubscript{6} system. Spin-polarized TDOS and projected density of state (PDOS) for strain-free (b)i-(f)I and strained (b)ii-(f)ii systems respectively.}
\label{fig4}
\end{figure}

\begin{figure}[!h]
\begin{center}
\includegraphics[width=6.2291in,height=3.04in]{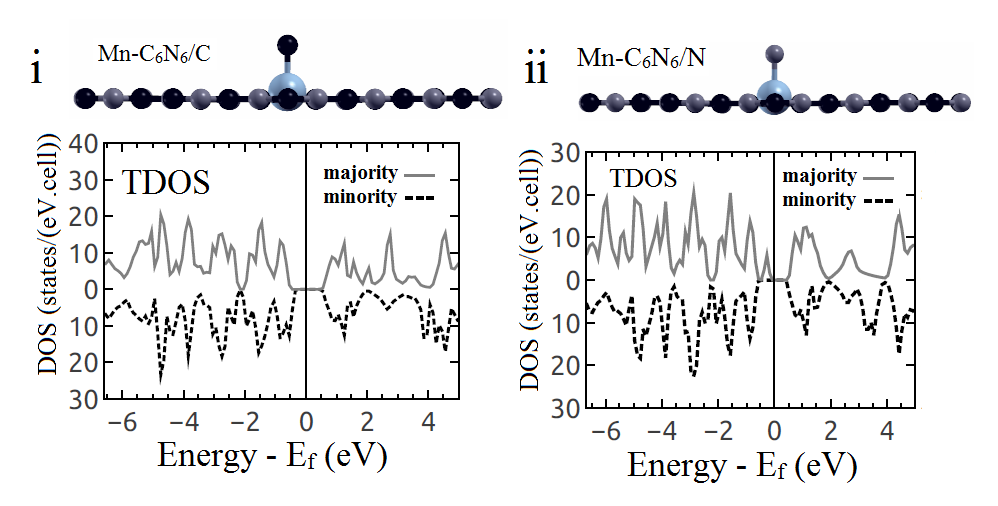}
\includegraphics[width=6.2291in,height=3.04in]{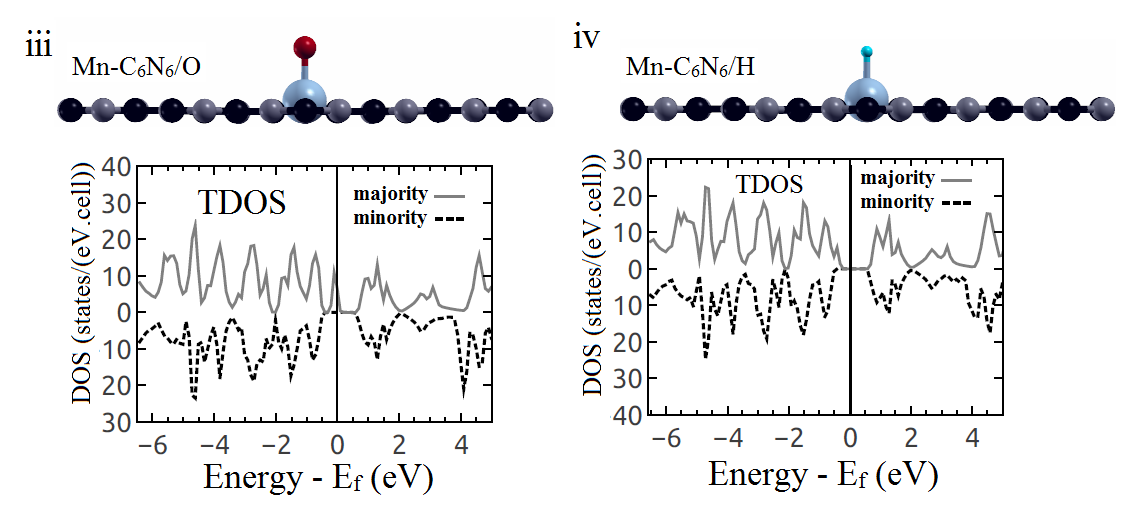}
\end{center}
\caption{Spin-polarized TDOS and side view for Mn-C\textsubscript{6}N\textsubscript{6} with an adsorbed (i) C
(ii) N (iii) O and (iv) H atoms. }
\label{fig5}
\end{figure}

\begin{figure}[!h]
{\centering  \includegraphics[width=6.2291in,height=3.04in]{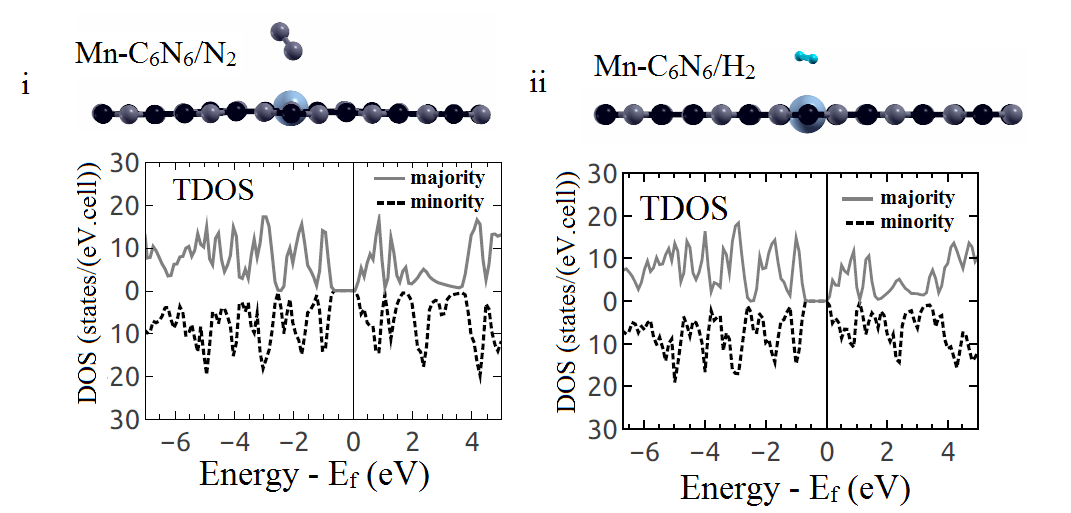} \par}
{\centering  \includegraphics[width=6.2291in,height=3.04in]{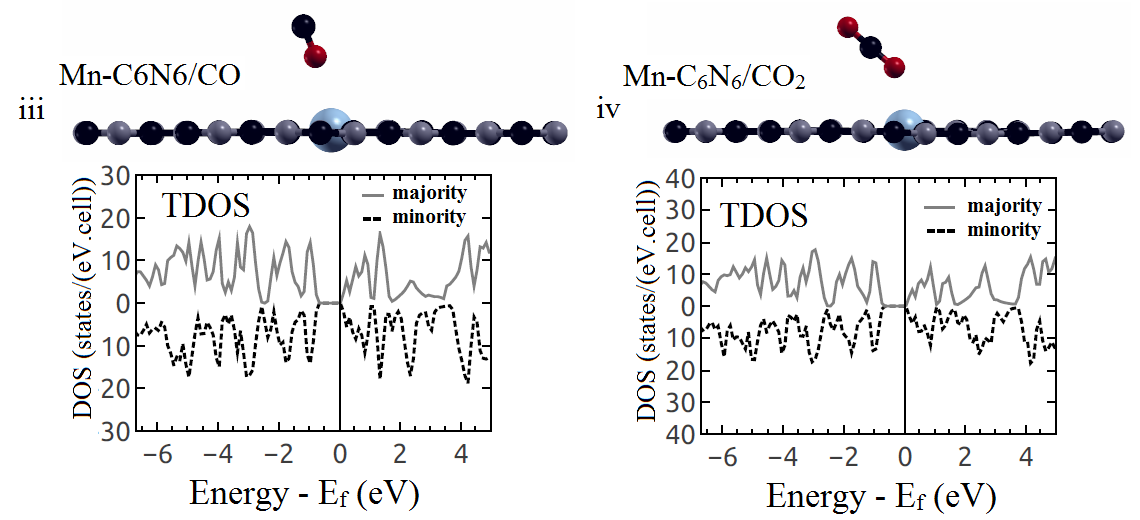} \par}
{\centering  \includegraphics[width=6.2291in,height=3.04in]{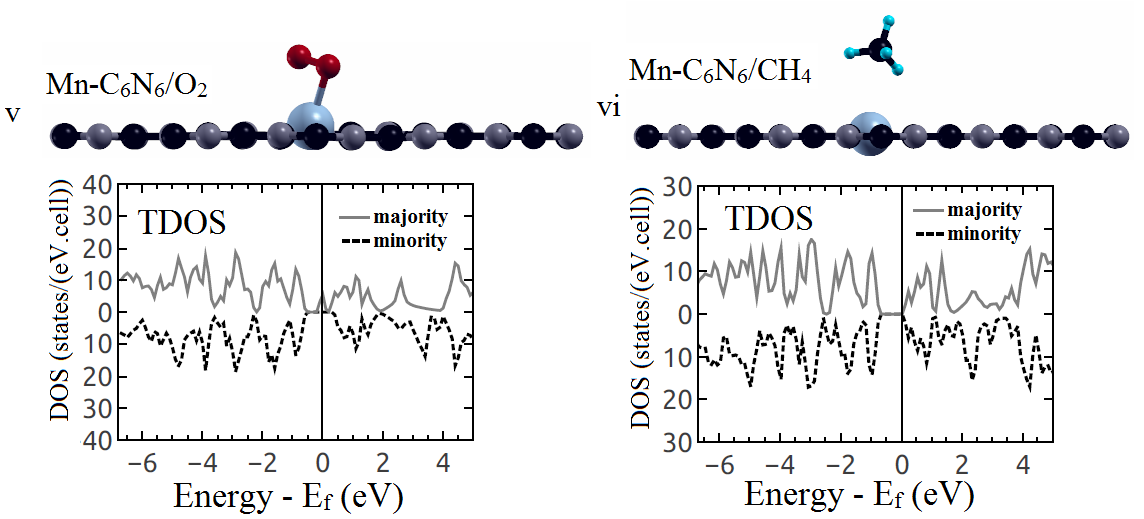} \par}
\caption{Spin-polarized TDOS and side view for Mn-C\textsubscript{6}N\textsubscript{6} with an adsorbed (i)
N\textsubscript{2} (ii) H\textsubscript{2} (iii) CO (iv) CO\textsubscript{2} (v) O\textsubscript{2} and (vi)
CH\textsubscript{4} molecules. }
\label{fig6}
\end{figure}

\end{document}